\documentclass[11pt,a4paper]{article}
\usepackage{amsmath,amssymb}
\usepackage{epsfig,graphicx}
\usepackage{cite}

\begin{document}

\title{\bf The third type of fermion mixing and indirect limits on the Pati--Salam leptoquark mass} 

\author{A.~V.~Kuznetsov\footnote{{\bf e-mail}: avkuzn@uniyar.ac.ru},
N.~V.~Mikheev\footnote{{\bf e-mail}: mikheev@uniyar.ac.ru},
A.~V.~Serghienko\footnote{{\bf e-mail}: serghienko@gmail.com}
\\
\small{\em Yaroslavl State P.G.~Demidov University} \\
\small{\em Sovietskaya 14, 150000 Yaroslavl, Russian Federation}
}
\date{}

\maketitle

\begin{abstract}
The low-energy manifestations of a minimal extension of the 
electroweak standard model based on the quark-lepton symmetry $SU(4)_V \otimes SU(2)_L
\otimes G_R$ of the Pati--Salam type are analyzed. Given this
symmetry the third type of mixing in the interactions of the
$SU(4)_V$ leptoquarks with quarks and leptons is shown to be
required. An additional arbitrariness of the mixing parameters could
allow, in principle, to decrease noticeably the indirect lower bound on the
vector leptoquark mass originated from the low-energy rare
processes, strongly suppressed in the standard model.
\end{abstract}

\section{Introduction}
\label{sec:Introduction}

\def\D{\mathrm{d}} 
\def\E{\mathrm{e}}
\def\I{\mathrm{i}}

While the LHC methodically examines the energy scale of the electroweak theory and above, 
it is time to recall the two criteria for evaluating a physical theory, 
mentioned by A. Einstein~\cite{Einstein:1949}.
The first point of view is obvious: a theory must not contradict empirical facts, 
and it is called the ``external confirmation''. 
The test of this criterion both for the standard model and its various extensions is now engaged in the LHC.  
The second point of view called the ``inner perfection'' of the theory, 
may be very important to refine the search area for new physics.

All existing experimental data in particle physics are in good agreement with 
the standard model predictions. However, the problems exist which could 
not be resolved within the standard model and it is obviously not a complete or final 
theory. It is unquestionable that the standard model should be the low-energy limit of 
some higher symmetry. The question is what could be this symmetry. And 
the main question is, what is the mass scale of this symmetry restoration. 
A gloomy prospect is the restoration of this higher symmetry at once on a 
very high mass scale, the so-called gauge desert. A concept of a consecutive 
symmetry restoration is much more attractive. It looks natural in this case to 
suppose a correspondence of the hierarchy of symmetries and the hierarchy 
of the mass scales of their restoration. Now we are on the first step of some 
stairway of symmetries and we try to guess what could be the next one. If 
we consider some well-known higher symmetries from this point of view, two 
questions are pertinent. First, isn't the supersymmetry\cite{Nilles:1984} as the symmetry of 
bosons and fermions, higher than the symmetry within the fermion sector, 
namely, the quark-lepton symmetry~\cite{Pati:1974}, or the symmetry within the boson 
sector, namely, the left-right 
symmetry~\cite{Lipmanov:1967,Lipmanov:1968a,Lipmanov:1968,Beg:1977}? 
Second, wouldn't the supersymmetry 
restoration be connected with a higher mass scale than the others? 
The recent searches for supersymmetry carried out
at the Tevatron and the LHC colliders~\cite{Portell_Bueso:2011} 
shown that no significant deviations from the
standard model predictions have been found, the vast parameter
space available for supersymmetry has been substantially reduced
and the most probable scenarios predicted by
electroweak precision tests are now excluded or under
some constraints after the new stringent limits.

We should like to analyse a possibility when the quark-lepton symmetry 
is the next step beyond the standard model. 
Along with the ``inner perfection'' argument for this theory, 
there exists a direct evidence in favor of it. 
The puzzle of fermion generations is recognized as one of the most 
outstanding problems of present particle physics, and may be the main justification for the need 
to go beyond the standard model. Namely, the cancellation of triangle axial anomalies which is 
necessary for the standard model to be renormalized, requires that fermions be grouped into generations.
This association provides an equation $\sum_f \, T_{3f} \, Q_f^2 = 0$, where the summation is taken over all 
fermions of a generation, both quarks of three colors and leptons, 
$T_{3f}$ is the 3d component of the weak isospin, and $Q_f$ is the electric charge of a fermion. 
Due to this equation, the divergent axial-vector part of the triangle $Z \gamma \gamma$ diagram 
with a fermion loop vanishes. 

The model where a combination of quarks and leptons into generations looked the most natural, 
proposed by J.C. Pati and A. Salam~\cite{Pati:1974} was based on the 
quark-lepton symmetry. The lepton number was treated in the model as the fourth color. 
As the minimal gauge group realizing this symmetry, one can consider the semi-simple group $SU(4)_V \otimes SU(2)_L \otimes G_R$. To begin with, one can take the group $U(1)_R$ as $G_R$. 
The fermions were combined into the fundamental representations of the 
$SU(4)_V$ subgroup, the neutrinos with the \emph{up} quarks and the charged leptons 
with the \emph{down} quarks: 
\begin{equation}
\left ( \begin{array}{c} u^1 \\ u^2 \\ u^3 \\ \nu \end{array}
\right )_i \, , \qquad \left (
\begin{array}{c} d^1 \\ d^2 \\ d^3 \\ \ell \end{array} \right )_i \, ,
\qquad i=1,2,3 \dots \, (?) \,, 
\label{eq:q}
\end{equation}
where the superscripts 1,2,3 number colors and the subscript $i$ numbers fermion generations, 
i.e. $u_i$ denotes $u, c, t, \dots$ and $d_i$ denotes $d, s, b, \dots$.

The left-handed fermions form fundamental representations of the
$SU(2)_L$ subgroup:
\begin{equation}
\left ( \begin{array}{c} u^c \\ d^c \end{array}
\right )_L \, , \qquad 
\left ( \begin{array}{c} \nu \\ \ell \end{array} \right )_L \, .
\label{eq:d}
\end{equation}
One should keep in mind that  
when considering the mass eigenstates, it is necessary to take into 
account the mixing of fermion states~\eqref{eq:q}, \eqref{eq:d}, 
to be analysed below.  
 
Let us remind that such an extension of
the standard model has a number of attractive features. 

\begin{enumerate}
\item
As it was mentioned above, definite quark-lepton symmetry is necessary in order that the
standard model be renormalized: cancellation of triangle anomalies
requires that fermions be grouped into generations.

\item
There is no proton decay because the lepton charge treated as the
fourth color is strictly conserved.

\item
Rigid assignment of quarks and leptons to representations~\eqref{eq:q} leads to 
a natural explanation for a fractional quark hypercharge. Indeed, 
the traceless 15-th generator $T_{15}^V$ of the $SU(4)_V$ subgroup can be represented in
the form 
\begin{equation} 
T_{15}^V=\sqrt{\frac{3}{8}} \; \text{diag}\left(\frac{1}{3}\,,\,\frac{1}{3}\,,\,\frac{1}{3}\,,\,-1\right) 
=\sqrt{\frac{3}{8}} \; Y_V \,. 
\label{eq:T15}
\end{equation} 
It is remarkable that the values of the standard model hypercharge of the left-handed quarks and leptons 
combined into the $SU(2)_L$ doublets turn out to be placed on the diagonal. 
Let us call it the vector hypercharge, $Y_V$, and assume that it belongs to both 
the left- and right-handed fermions. 

\item
Let us suppose that $G_R = U(1)_R$. The well-known values 
of the standard model hypercharge of the left and right, 
and \emph{up} and \emph{down} quarks and 
leptons are:
\begin{equation}
Y_{SM} \; = \; \left \{
\begin{array}{c} \left (\begin{array}{c} \frac{1}{3} \\ \\ \frac{1}{3}
\end{array} \right ) \quad \mbox{for} \; q_L ; \\ \\
\left (\begin{array}{c} \frac{4}{3} \\ \\ -\frac{2}{3}
\end{array} \right ) \quad \mbox{for} \; q_R ; \end{array}
\begin{array}{c} \left (\begin{array}{c} - 1 \\ \\ - 1
\end{array} \right ) \quad \mbox{for} \; \ell_L \\ \\
\left (\begin{array}{c} 0 \\ \\ - 2
\end{array} \right ) \quad \mbox{for} \; \ell_R \end{array}
\right \}.
\label{eq:Y}
\end{equation}
Then, from the equation $Y_{SM} = Y_V + Y_R$, taking Eq.~\eqref{eq:T15} into account, 
one obtains that the values of the right 
hypercharge $Y_R$ occur to be equal $\pm 1$ for the \emph{up} and \emph{down} 
fermions correspondingly, both quarks and leptons. 
It is tempting to interpret this circumstance as the indication
that the right-hand hypercharge is the doubled third component of
the right-hand isospin. Thus, the subgroup $G_R$ may be $SU(2)_R$.
\end{enumerate}

``Under these circumstances one would be surprised if Nature had made no use of it'', 
as P.~Dirac wrote on another occasion~\cite{Dirac:1931}. 

The most exotic object of the Pati--Salam type symmetry is the charged 
and colored gauge $X$ boson named leptoquark. Its mass $M_X$ should be the 
scale of breaking of $SU(4)_V$ to $SU(3)_c$. 
Bounds on the vector leptoquark mass are obtained both directly and
indirectly, see Ref.~\cite{Beringer:2012}.
The direct search\cite{Aaltonen:2008_LQ} for vector leptoquarks using $\tau^+ \tau^- b \bar b$ 
events in $p \bar p$ collisions at $E_{cm} = 1.96$ TeV have provided the lower mass limit at a level 
of 250--300 GeV, depending on the coupling assumed.  
Much more stringent indirect limits are calculated from the bounds
on the leptoquark-induced four-fermion interactions, which are
obtained from low-energy experiments. There is an extensive series of papers where 
such indirect limits on the vector leptoquark mass were estimated, 
see e.g. Refs.~\cite{Shanker:1982,Deshpande:1983,Leurer:1994b,Davidson:1994,Valencia:1994,Kuznetsov:1994,
Kuznetsov:1995,Smirnov:1995a,Smirnov:1995b,Smirnov:2007,Smirnov:2008}. 
The most stringent bounds\cite{Beringer:2012} 
were obtained from the data on the $\pi \to e \nu$ decay and 
from the upper limits on the $K_L^0 \to e \mu$ and $B^0 \to e \tau$ decays. 
However, those estimations were not comprehensive 
because the phenomenon of a mixing in the lepton-quark currents was not 
considered there. It will be shown that such a mixing inevitably occurs in 
the theory. 

An important part of the model under consideration is its scalar sector, which 
also contains exotic objects such as scalar leptoquarks. We do not concern here 
the scalar sector, which could be much more ambiguous than the gauge one. Such an analysis can be 
found e.g. in Refs.~\cite{Smirnov:2007,Smirnov:2008,Leurer:1994a}.

The paper is organized as follows. 
In Sec.~\ref{sec:Mixing}, it is argued that three types of fermion mixing 
inevitably arise at the loop level if initially fermions are taken without mixing. 
The effective four-fermion Lagrangian caused by the leptoquark interactions with 
quarks and leptons is presented in Sec.~\ref{sec:Lagrangian}. 
In Sec.~\ref{sec:Constraints}, we update the constraints on the parameters of the scheme 
which were obtained in our recent paper~\cite{Kuznetsov:2012} on a base of the data 
from different low-energy processes which are 
strongly suppressed or forbidden in the standard model. 
The updating of the constraint on the vector leptoquark mass is made in 
Sec.~\ref{sec:Update} basing on a new data from 
CMS and LHCb Collaborations on the rare decays 
$B^0_{d,s} \to \mu^+ \mu^-$~\cite{CMS:2012,LHCb:2012,Gushchin:2012}.

\section{The third type of fermion mixing}
\label{sec:Mixing}

As the result of the Higgs mechanism in the Pati--Salam model,
fractionally charged colored gauge $X$-bosons, vector leptoquarks appear. 
Leptoquarks are responsible for transitions between
quarks and leptons. The scale of the breakdown of $SU(4)_V$ symmetry
to $SU(3)_c$ is the leptoquark mass $M_X$. The three fermion
generations are grouped into the following $\{4,2\}$ representations
of the $SU(4)_V\otimes SU(2)_L$ group:
\begin{equation}
\begin{pmatrix} u^c & d^c\\
\nu & \ell
\end{pmatrix}_i~\left(i=1,2,3\right).
\label{eq:mixing}
\end{equation}
where $c$ is the color index to be further omitted. 
It is known that there exists the mixing of quarks
in weak charged currents, which is described by the
Cabibbo-Kobayashi-Maskawa matrix. Therefore, at least one of the
states in \eqref{eq:mixing}, $u$ or $d$, is not diagonal in mass. It can easily be seen
that, because of mixing that arises at the loop level, none of the
components is generally a mass eigenstate. As usual, we assume that
all the states in \eqref{eq:mixing}, with the exception of $d$, are initially diagonal in
mass. This leads to nondiagonal transitions $\ell \to X + d (s,b) \to \ell^\prime$
through a quark-leptoquark loop, see Fig.~1. As this diagram is divergent, the corresponding 
counterterm should exist at the tree level. This means that the lepton states $\ell$ 
in \eqref{eq:mixing} are not the mass eigenstates, and there is mixing in the
lepton sector. Other nondiagonal transitions arise in a similar way.
Hence, in order that the theory be renormalizable, it is necessary
to introduce all kinds of mixing even at the tree level.
%
\begin{figure}
\begin{center}
\includegraphics*[width=0.3\textwidth]{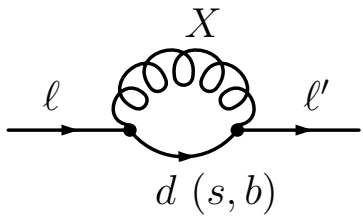}
\caption{Feynman diagram illustrating the appearance of fermion mixings.} 
\end{center}
\label{fig:X_loop}
\end{figure}
%
As all the fermion representations are identical, they can always be
regrouped in such a way that one state is diagonal in mass. The most
natural way is to diagonalize charged leptons. In this case, fermion
representations can be written in the form
\begin{equation} 
\begin{pmatrix} u & d\\
\nu & \ell
\end{pmatrix}_{\ell}=
\begin{pmatrix} u_e & d_e\\
\nu_e & e
\end{pmatrix},~
\begin{pmatrix} u_\mu & d_\mu\\
\nu_\mu & \mu
\end{pmatrix},~
\begin{pmatrix} u_\tau & d_\tau\\
\nu_\tau & \tau
\end{pmatrix}.
\label{eq:repr1} 
\end{equation}
Here, the quarks and neutrinos subscripts $\ell=e,\mu,\tau$ 
label the states which are not mass eigenstates and which enter into the
same representation as the charged lepton $\ell$: 
\begin{equation} 
\nu_{\ell}=\sum_i {\cal K}_{{\ell} i}\nu_i \,, \quad 
u_{\ell}=\sum_p {\cal U}_{{\ell} p}u_p\,, \quad 
d_{\ell}=\sum_n {\cal D}_{{\ell} n}d_n \,. 
\label{eq:repr2} 
\end{equation}
Here, ${\cal K}_{{\ell}i}$ is the unitary leptonic mixing matrix 
by Pontecorvo--Maki--Nakagawa--Sakata.
The matrices ${\cal U}_{{\ell}p}$ and ${\cal D}_{{\ell}n}$ are the
unitary mixing matrices in the interactions of leptoquarks with the \emph{up} and \emph{down} 
fermions correspondingly, both quarks and leptons. 
The states $\nu_i,~u_p$ and $d_n$ are the mass eigenstates: 
\begin{equation} \begin{aligned} &\nu_i=\left(\nu_1,\nu_2,\nu_3\right),\\
&u_p=\left(u_1,u_2,u_3\right)=\left(u,c,t\right),\\
&d_n=\left(d_1,d_2,d_3\right)=\left(d,s,b\right). \end{aligned} 
\label{eq:repr3} 
\end{equation} 
Thus, there are generally three types of mixing in this scheme.

In our notation, the well-known Lagrangian describing the
interaction of charge weak currents with $W$-bosons takes the form
\begin{eqnarray} 
{\cal L}_W &=& \frac{g}{2\sqrt{2}}\left[\left(\bar{\nu}_{\ell} O_\alpha \ell \right)+\left(\bar{u}_{\ell} O_\alpha d_{\ell} \right)\right] W^\dagger_\alpha + \text{h.c.}
\nonumber\\
&=&\frac{g}{2\sqrt{2}}\left[{\cal K}^*_{\ell i}\left(\bar{\nu}_i O_\alpha \ell \right) 
+ {\cal U}^*_{\ell p} {\cal D}_{\ell n} \left(\bar{u}_p O_\alpha d_n \right)\right]W^\dagger_\alpha + \text{h.c.},
\label{eq:Lagr_W} 
\end{eqnarray} 
where $g$ is the constant of the $SU(2)_L$ group and
$O_\alpha=\gamma_\alpha\left(1-\gamma_5\right)$. It follows that the standard
Cabibbo--Kobayashi--Maskawa matrix is $V={\cal U}^\dagger {\cal D}$. This is
the only available information about the matrices ${\cal U}$ and
${\cal D}$ of mixing in the leptoquark sector. The matrix ${\cal K}$
describing a mixing in the lepton sector is the subject of intensive
experimental studies.

Following the spontaneous breakdown of the $SU(4)_V$ symmetry to $SU(3)_c$
on the scale of $M_X$, six massive vector bosons forming three charged
colored leptoquarks, decouple from the 15-plet of gauge fields. The
interaction of these leptoquarks with fermions has the form 
\begin{equation}
{\cal L}_X=\frac{g_S\left(M_X\right)}{\sqrt{2}} \left[ {\cal D}_{\ell n} \left(\bar{\ell} \gamma_\alpha d^c_n\right) + 
\left({\cal K}^\dagger {\cal U}\right)_{ip} \left(\bar{\nu}_i \gamma_\alpha u^c_p \right) \right] X^c_\alpha + \text{h.c.},
\label{eq:LagDen} 
\end{equation} 
where the color superscript $c$ is written
explicitly once again. The coupling constant $g_S\left(M_X\right)$ is
expressed in terms of the strong-interaction constant $\alpha_S$ on the
scale of the leptoquark mass $M_X$ as
$g_S^2\left(M_X\right)/4\pi=\alpha_S\left(M_X\right)$.

\section{Effective Lagrangian with allowance for QCD corrections}
\label{sec:Lagrangian}

If the momentum transfer satisfies the condition $q^2\ll M_X^2$,
the Lagrangian \eqref{eq:LagDen} leads to the effective four-fermion
vector-vector interaction between quarks and leptons. By applying
the Fierz transformation, we can isolate the lepton and quark currents
(scalar, pseudoscalar, vector and axial-vector currents) in the
effective Lagrangian. In constructing the effective Lagrangian of
leptoquark interactions, it is necessary to take into account the QCD
corrections, which can easily be estimated, see e.g. Refs.~\cite{Vainstein,Vysotskii}. 
In the case under study, we
can use the approximation of leading logarithms because
$\ln\left(M_X/\mu\right)\gg1$, where $\mu\sim1~\text{GeV}$ is the typical
hadronic scale. As the result of taking the QCD corrections into
account, the scalar and pseudoscalar coupling constants acquire the
enhancement factor 
\begin{equation}
Q\left(\mu\right) = \left(\frac{\alpha_S\left(\mu\right)}{\alpha_S\left(M_X\right)}\right)^{4/\bar{b}},
\label{eq:EnFact} 
\end{equation} 
where $\alpha_S\left(\mu\right)$ is the strong-interaction
constant on the scale $\mu$, $\bar{b}=11-2/3\left(\bar{n}_f\right)$, and
$\bar{n}_f$ is the mean number of quark flavors on the scales
$\mu^2\leq q^2\leq M_X^2$; for $M_X^2\gg m_t^2$, we have
$\bar{b}\simeq7$.

Further we investigate the contribution to low-energy processes from the 
interaction Lagrangian \eqref{eq:LagDen} involving leptoquarks and find
constraints on the parameters of the scheme from available
experimental data. As the analysis shows, the most stringent
constraints on the vector-leptoquark mass $M_X$ and on the elements
of the mixing matrix ${\cal D}$ follow from the data on rare 
$\pi$ and $K$ meson decays. 

Possible constraints on the masses and coupling constants of vector
leptoquarks from experimental data on rare $\pi$ and $K$ decays were
analyzed in Refs.~\cite{Shanker:1982,Deshpande:1983,Leurer:1994b,Davidson:1994,Valencia:1994,Kuznetsov:1994,
Kuznetsov:1995,Smirnov:1995a,Smirnov:1995b,Smirnov:2007,Smirnov:2008}. 
One approach\cite{Shanker:1982,Leurer:1994b,Davidson:1994} was based on using 
the phenomenological model-independent Lagrangians describing the interactions of
leptoquarks with quarks and leptons. 
Pati--Salam quark-lepton symmetry was considered in
Refs.~\cite{Deshpande:1983,Valencia:1994,Kuznetsov:1994,Kuznetsov:1995,Smirnov:1995a,
Smirnov:1995b,Smirnov:2007,Smirnov:2008}. 
QCD corrections were included into an analysis in Refs.~\cite{Valencia:1994,Kuznetsov:1994,Kuznetsov:1995}. 
The authors of Ref.~\cite{Valencia:1994} considered
the possibility of mixing in quark-lepton currents, but they
analyzed only specific cases in which each charged lepton is
associated with one quark generation. In our notation, this
corresponds to the matrices ${\cal D}$ that are obtained from the
unit matrix by making all possible permutation of columns.

In the description of the $\pi$- and $K$-meson interactions, it is
sufficient to retain only the scalar and pseudoscalar coupling
constants in the effective Lagrangian. Really, these couplings are more
significant in the amplitudes, because they are enhanced, first, by QCD corrections, 
and second, by the smallness of the current-quark masses arising in 
the amplitude denominators. 
The corresponding part of the effective Lagrangian can be represented as 
\begin{eqnarray} 
\Delta{\cal L}_{\pi,K} &=& -\frac{2\pi\alpha_S\left(M_X\right)}{M_X^2} \, Q \left(\mu\right) 
\left[{\cal D}_{\ell n} \left( {\cal U}^\dagger {\cal K}\right)_{pi} 
\left(\bar{\ell} \gamma_5 \nu_i \right) \left(\bar{u}_p\gamma_5d_n\right) + \text{h.c.} 
- \left( \gamma_5 \to 1 \right)\right]
\nonumber\\
&-&\frac{2\pi\alpha_S\left(M_X\right)}{M_X^2} \, Q \left(\mu\right) 
\bigg[{\cal D}_{\ell n} {\cal D}^*_{\ell^\prime n^\prime} 
\left(\bar{\ell} \gamma_5 \ell^\prime \right) \left( \bar{d}_{n^\prime} \gamma_5 d_n \right)  
\nonumber\\
&+& \left({\cal K}^\dagger {\cal U}\right)_{ip}\left({\cal U}^\dagger {\cal K}\right)_{p^\prime
i^\prime}\left(\bar{\nu}_i\gamma_5\nu_{i^\prime}\right)\left(\bar{u}_{p^\prime}\gamma_5u_p\right)-\left(\gamma_5\to1\right)
\bigg].
\label{eq:LagPiK}  
\end{eqnarray} 
This Lagrangian contributes to the rare $\pi$, $K$, $\tau$ and $B$ decays, 
which are strongly suppressed or forbidden in the standard model.
For the $\tau$ and $B$ decays, this Lagrangian is not enough, and a part 
with the product of axial-vector currents should be added. 

\section{Constraints on the parameters of the scheme from low-energy processes}
\label{sec:Constraints}

In our recent paper~\cite{Kuznetsov:2012}, we have performed a detailed analysis of a large set 
of experimental data on different low-energy processes which are 
strongly suppressed or forbidden in the standard model. The constraints on the vector leptoquark mass 
were obtained. In Table~\ref{tab:1}, the most stringent constraints of Ref.~\cite{Kuznetsov:2012} 
are summarized. All the constraints involve the elements of the unknown unitary mixing matrix ${\cal D}$:
\begin{equation} 
{\cal D}_{\ell n} = 
\begin{pmatrix} 
{\cal D}_{e d} & {\cal D}_{e s} & {\cal D}_{e b}\\[2mm]
{\cal D}_{\mu d} & {\cal D}_{\mu s} & {\cal D}_{\mu b}\\[2mm]
{\cal D}_{\tau d} & {\cal D}_{\tau s} & {\cal D}_{\tau b}
\end{pmatrix}.
\label{Ddef} 
\end{equation}

The possibility was analysed in Ref.~\cite{Kuznetsov:2012} for the constraints 
on the vector leptoquark mass $M_X$ to be much weaker than the numbers 
in Table~\ref{tab:1}. The case was considered when the 
elements ${\cal D}_{ed}$ and ${\cal D}_{es}$ 
are small enough, to eliminate the most strong restriction arising from the  
limit on the decays $K^0_L \to e^\pm \mu^\mp$. For evaluation, these elements were 
taken to be zero. 
Given the unitarity of the matrix ${\cal D}$, this meant that ${\cal D}_{eb} = 1$, 
and ${\cal D}_{\mu b} = {\cal D}_{\tau b} = 0$. 
The remaining $(2 \times 2)$-matrix was parameterized by one angle. 
The insertion of the phase factor allowed to eliminate the restriction arising from the  
limit on $Br(K^0_L \to \mu^+ \mu^-)$ which contained the real part of the 
${\cal D}$ matrix elements product. The ${\cal D}$ matrix was taken in the form:
\begin{equation} 
{\cal D}_{\ell n} \simeq 
\begin{pmatrix} 
0 & 0 & 1~\\[2mm]
\cos \varphi & ~\text{i} \sin \varphi~  & 0~\\[2mm]
~\text{i} \sin \varphi & \cos \varphi & 0~
\end{pmatrix}.
\label{eq:Dfin} 
\end{equation}
%

\begin{table}[ht]
\caption{Constraints on the leptoquark mass and on the elements of
the ${\cal D}$ matrix from experimental data on rare decays.}
\begin{center}
{\begin{tabular}{lcl} 
\\
\hline
\\
Experimental limit & Ref. & Bound 
\\ \\
\hline 
\\
\bigskip
$Br(K^0_L \to e^\pm \mu^\mp) < 4.7 \times 10^{-12}$ 
& \cite{AMBROSE98B} & $\frac{\mbox{\footnotesize $M_X$}}
{\mbox{\footnotesize $|{\cal D}_{ed} {\cal D}^*_{\mu s}+{\cal D}_{es}
{\cal D}^*_{\mu d}|^{1/2}$}} \,
> \, 2100~\textrm{TeV}$ \\
\bigskip
$Br(K^0_L \to \mu^+ \mu^-) = (6.84\pm0.11) \times 10^{-9}$ 
& \cite{AMBROSE00,ALEXOPOULOS04} 
& $\frac{\mbox{\footnotesize
$M_X$}} {\mbox{\footnotesize $|\text{Re}({\cal D}_{\mu d} {\cal D}^*_{\mu
s})|^{1/2}$}} \,
> \, 1100~\textrm{TeV}$ \\
\bigskip
$Br(B^0\to e^+ \mu^-)<6.4\times10^{-8}$  
& \cite{Aaltonen:2009} 
& $\frac{\mbox{\footnotesize $M_X$}}
{\mbox{\footnotesize $|{\cal D}_{\mu d} {\cal D}_{e b}|^{1/2}$}} \,
> \, 55~\textrm{TeV}$ \\
\bigskip
$Br(B^0\to\mu^+\mu^-)<1.5\times10^{-8}$ 
& \cite{Aaltonen:2008} & $\frac{\mbox{\footnotesize $M_X$}}
{\mbox{\footnotesize $|{\cal D}_{\mu d} {\cal D}_{\mu b}|^{1/2}$}} \,
> \, 79~\textrm{TeV}$ \\
\bigskip
$Br(B^0_s\to e^+ \mu^-)<2.0\times10^{-7}$ 
& \cite{Aaltonen:2009} & $\frac{\mbox{\footnotesize $M_X$}}
{\mbox{\footnotesize $|{\cal D}_{\mu s} {\cal D}_{e b}|^{1/2}$}} \,
> \, 41~\textrm{TeV}$ \\
\bigskip
$Br(B^0_s\to\mu^+\mu^-)<4.2\times10^{-8}$ 
& \cite{Abazov:2010} & $\frac{\mbox{\footnotesize $M_X$}}
{\mbox{\footnotesize $|{\cal D}_{\mu s} {\cal D}_{\mu b}|^{1/2}$}} \,
> \, 61~\textrm{TeV}$ \\
\hline
\end{tabular}
\label{tab:1}}
\end{center}
\end{table}

The constraints on the vector leptoquark mass and the $\varphi$ angle 
arising from Table~\ref{tab:1} took the form:
\newline
i) $B^0\to e^+ \mu^-$
\begin{equation} 
M_X > 55~\textrm{TeV} \, |\cos \varphi|^{1/2} \,,
\label{fin3} 
\end{equation}
i1) $B^0_s\to e^+ \mu^-$
\begin{equation} 
M_X > 41~\textrm{TeV} \, |\sin \varphi|^{1/2} \,.
\label{fin4} 
\end{equation}
Combining these constraints, the limit on the vector leptoquark mass 
was obtained~\cite{Kuznetsov:2012}: 
\begin{equation} 
M_X > 38~\textrm{TeV} \,.
\label{eq:finMX} 
\end{equation}
%

\section{Different mixings for left-handed and right-handed fermions}
\label{sec:Different}

We have considered a possibility when the quark-lepton symmetry 
was the next step beyond the standard model. Then the left-right symmetry which is 
believed to exist in Nature, should restore at higher mass scale. 
But this means that the left-right symmetry should be already broken at the scale $M_X$. 
It is worthwhile to consider the matrices ${\cal D}^{(L)}, {\cal U}^{(L)}$ and 
${\cal D}^{(R)}, {\cal U}^{(R)}$ which are in a general case different 
for left-handed and right-handed fermions. 
This possibility and some its consequences were also considered 
in Refs.~\cite{Smirnov:1995a,Smirnov:1995b,Smirnov:2007,Smirnov:2008}.
The interaction Lagrangian of leptoquarks with fermions takes the form instead of 
Eq.~\eqref{eq:LagDen}:
\begin{eqnarray}
{\cal L}_X &=& \frac{g_S\left(M_X\right)}{2 \sqrt{2}} \bigg[ 
{\cal D}^{(L)}_{\ell n} \left(\bar{\ell} O_\alpha d_n \right) + 
{\cal D}^{(R)}_{\ell n} \left(\bar{\ell} O_\alpha^\prime d_n \right) 
\nonumber\\
&+& \left( {\cal K}^{(L)\dagger} {\cal U}^{(L)} \right)_{ip} \left(\bar{\nu}_i O_\alpha u_p \right) 
+ \left( {\cal K}^{(R)\dagger} {\cal U}^{(R)} \right)_{ip} \left(\bar{\nu}_i O_\alpha^\prime u_p \right) 
\bigg] 
X_\alpha + \text{h.c.},
\label{Lagr_LR} 
\end{eqnarray} 
where $O_\alpha=\gamma_\alpha\left(1-\gamma_5\right)$, $O_\alpha^\prime=\gamma_\alpha\left(1+\gamma_5\right)$. 

The constraints on the model parameters from experimental data on rare $\pi$ and $K$ decays  
in the case of different mixings take the forms presented in Table 5 of Ref.~\cite{Kuznetsov:2012}. 
If one would wish to reduce the limits on $M_X$ presented there from thousands and hundreds 
to tens of TeV by varying the elements of the ${\cal D}^{(L)}$ and ${\cal D}^{(R)}$ matrices, 
it seems that the elements ${\cal D}^{(L)}_{e d}$ and ${\cal D}^{(R)}_{e d}$ should be taken small in any case.  
If one takes them for evaluation be zero, the most strong restriction from 
the limit on $Br(K^0_L \to e^\pm \mu^\mp)$ acquires the form:
\begin{equation} 
\frac{M_X}
{\left(\left|{\cal D}^{(L)}_{es} {\cal D}^{(R)}_{\mu d}\right|^2 
+ \left|{\cal D}^{(R)}_{es} {\cal D}^{(L)}_{\mu d}\right|^2 
\right)^{1/4}} > 1770~\textrm{TeV} \,.
\label{LR_2} 
\end{equation}

There are two possibilities to eliminate this bound, 
which we call the symmetric and the asymmetric cases.

\emph{The symmetric case} is realized when both of the matrices ${\cal D}^{(L)}$ 
and ${\cal D}^{(R)}$ are taken in the form of Eq.~\eqref{eq:Dfin} with the angles $\varphi_L$ and $\varphi_R$. 
In this case the restriction from the limit on $Br(K^0_L \to \mu^+ \mu^-)$ 
takes the form:
\begin{equation} 
M_X > 780~\textrm{TeV} \, |\sin \left( \varphi_L - \varphi_R \right)|^{1/2} \,.
\label{LR_3} 
\end{equation}
To eliminate this bound, the angles should be close to each other or differ by $\pi$, in any case 
we come back to the result~\eqref{eq:finMX}. 

\emph{The asymmetric case} is realized when the matrices are taken in the form:
\begin{equation} 
{\cal D}^{(L)}_{\ell n} \simeq 
\begin{pmatrix} 
~0 & \cos \chi_L &  ~\sin \chi_L~\\[2mm]
~0 & - \sin \chi_L & ~\cos \chi_L~\\[2mm]
~1 & 0 & 0
\end{pmatrix} ,
\quad
{\cal D}^{(R)}_{\ell n} \simeq 
\begin{pmatrix} 
~0~~ & 0~~ & 1~~\\[2mm]
~0~~ & 1~~ & 0~~\\[2mm]
~1~~ & 0~~ & 0~~
\end{pmatrix} .
\label{LR_4} 
\end{equation}
As the analysis shows~\cite{Kuznetsov:2012}, the most stringent constraints arise 
from the following limits on the branching ratios of the processes:
\newline
i) $B^0_s \to \mu^+ \mu^-$
\begin{equation} 
M_X > 51~\textrm{TeV} \, |\cos \chi_L|^{1/2} \,,
\label{LR_6} 
\end{equation}
ii) $B^0_s \to e^+ \mu^-$
\begin{equation} 
M_X > 41~\textrm{TeV} \, |\sin \chi_L|^{1/2} \,.
\label{LR_7} 
\end{equation}
From these constraints, the limit was obtained~\cite{Kuznetsov:2012} on the vector leptoquark mass 
from low-energy processes in the case of different mixing matrices for left-handed and right-handed 
fermions, which coincided, with a good accuracy, with the limit \eqref{eq:finMX} obtained 
in the left-right-symmetric case: 
\begin{equation} 
M_X > 38~\textrm{TeV} \,.
\label{LR_8} 
\end{equation}
%

\section{Updated constraints from the LHC data}
\label{sec:Update}

The updating of the constraint on the vector leptoquark mass is based on a new data from 
CMS and LHCb Collaborations on the rare decays 
$B^0_{d,s} \to \mu^+ \mu^-$~\cite{CMS:2012,LHCb:2012,Gushchin:2012}, which are 
presented in Table~\ref{tab:2}.

\begin{table}[ht]
\caption{Constraints on the model parameters from new data of the  
CMS and LHCb Collaborations on the rare decays 
$B^0_{d,s} \to \mu^+ \mu^-$ (90 \% C.L.)}
\begin{center}
{\begin{tabular}{lcl} 
\\
\hline
\\
Experimental limit & Ref. & Bound 
\\ \\
\hline 
\\
\bigskip
$Br(B^0\to\mu^+\mu^-)<1.4\times10^{-9}$ 
& CMS~\cite{CMS:2012} & $\frac{\mbox{\footnotesize $M_X$}}
{\mbox{\footnotesize $|{\cal D}_{\mu d} {\cal D}_{\mu b}|^{1/2}$}} \,
> \, 143~\textrm{TeV}$ \\
\bigskip
$Br(B^0_s\to\mu^+\mu^-)<6.4\times10^{-9}$ 
& CMS~\cite{CMS:2012} & $\frac{\mbox{\footnotesize $M_X$}}
{\mbox{\footnotesize $|{\cal D}_{\mu s} {\cal D}_{\mu b}|^{1/2}$}} \,
> \, 98~\textrm{TeV}$ \\
\bigskip
$Br(B^0\to\mu^+\mu^-)<0.81\times10^{-9}$ 
& LHCb~\cite{LHCb:2012} & $\frac{\mbox{\footnotesize $M_X$}}
{\mbox{\footnotesize $|{\cal D}_{\mu d} {\cal D}_{\mu b}|^{1/2}$}} \,
> \, 164~\textrm{TeV}$ \\
\bigskip
$Br(B^0_s\to\mu^+\mu^-)<3.8\times10^{-9}$ 
& LHCb~\cite{LHCb:2012} & $\frac{\mbox{\footnotesize $M_X$}}
{\mbox{\footnotesize $|{\cal D}_{\mu s} {\cal D}_{\mu b}|^{1/2}$}} \,
> \, 112~\textrm{TeV}$ \\
\hline
\end{tabular}
\label{tab:2}}
\end{center}
\end{table}

These new data improve the constraints obtained in the asymmetric case~\eqref{LR_4}, 
namely, the data of the LHCb Collaboration on the decay $B^0_s \to \mu^+ \mu^-$ provide, 
instead of~\eqref{LR_6}:
\begin{equation} 
M_X > 94~\textrm{TeV} \, |\cos \chi_L|^{1/2} \,.
\label{LR_LHCb1} 
\end{equation}
Combining this bound with Eq.~\eqref{LR_7}, one obtains the final limit on the vector leptoquark mass 
in the case of different mixing matrices for left-handed and right-handed 
fermions: 
\begin{equation} 
M_X > 41~\textrm{TeV} \,.
\label{LR_LHCb2} 
\end{equation}
%

\section{Conclusion}

Thus, the detailed analysis of the available experimental data on rare
decays yields constraints on
the vector leptoquark mass that always involve the elements of the
unknown mixing matrix ${\cal D}$. 
Combining the most strong constraints from the experimental data on the low-energy processes, 
presented in Tables~\ref{tab:1} and~\ref{tab:2}, 
we have obtained in the case of identical mixings for left-handed and right-handed fermions
the following lowest limit on the vector leptoquark mass:  
$M_X > 38~\textrm{TeV}$. 
The lowest limit obtained in the asymmetric case~\eqref{LR_4} of different mixing matrices 
for left-handed and right-handed fermions appears to be: $M_X > 41~\textrm{TeV}$. 

\section*{Acknowledgements}

A.K. and N.M. express their deep gratitude to the organizers of the 
Seminar ``Quarks-2012'' for warm hospitality.  
We thank A.\,V.~Povarov and A.\,D.~Smirnov for useful discussions.
 
The study was performed within the State Assignment for Yaroslavl 
University (Project No.~2.4176.2011), and was supported in part by the 
Russian Foundation for Basic Research (Project No.~11-02-00394-a).



\end{document}